\documentclass[aps,twocolumn,superscriptaddress,altaffilletter,lengthcheck,tightenlines,showpacs,showkeys]{revtex4}

\newcommand{\al}{\alpha} 
\newcommand{\D}{\Delta} 
\newcommand{\ben}{\begin{eqnarray}} 
\newcommand{\een}{\end{eqnarray}} 
\newcommand{\be}{\begin{equation}} 
\newcommand{\ee}{\end{equation}} 
\newcommand{\ba}{\begin{eqnarray}} 
\newcommand{\ea}{\end{eqnarray}} 
\newcommand{\n}{\label}

\newcommand{\ga}{\gamma}

\usepackage[dvipdf]{epsfig}
\usepackage{color} 
\begin{document} 
 
\title{Cosmological model with interactions in the dark sector} 
\author{Luis P. Chimento}\email{chimento@df.uba.ar} 
\address{ Departamento de F\'{\i}sica, Facultad de Ciencias Exactas y 
Naturales,  Universidad de Buenos Aires, Ciudad 
Universitaria, Pabell\'on I, 1428 Buenos Aires, Argentina} 
\author{M\'onica Forte }\email{forte.monica@gmail.com} 
\address{ Departamento de F\'{\i}sica, Facultad de Ciencias Exactas y 
Naturales,  Universidad de Buenos Aires, Ciudad 
Universitaria, Pabell\'on I, 1428 Buenos Aires, Argentina} 
\author{Gilberto M. Kremer}\email{kremer@fisica.ufpr.br} 
\address{ Departamento de F\'\i sica, 
Universidade Federal do Paran\'a, Caixa Postal 19044, 81531-990 Curitiba, Brazil} 
 
\date{\today}

\begin{abstract} 
A cosmological model is proposed for the current Universe consisted of non-interacting baryonic matter and 
interacting dark components. The dark energy and dark matter are coupled through 
their effective barotropic indexes, which are considered as functions of the ratio between their energy densities. 
It is investigated two cases where the ratio is asymptotically stable and 
their parameters are adjusted by considering best fits to Hubble function data. It is shown that the deceleration parameter, the densities parameters,  and the luminosity distance have the correct behavior which is expected for a viable present scenario of the Universe. 
\end{abstract} 
 
\maketitle

\section{Introduction} 
 
 Recent observations of Supernova Ia (SNIa) suggest that the Universe has entered into a 
stage of an accelerated expansion with a redshift $z \lesssim 1$, 
\cite{Riess:1998cb}, \cite{Perlmutter:1998np}, \cite{Astier:2005qq}.  This has been 
confirmed by precise measurements of the spectrum of the Cosmic Microwave Background (CMB) 
anisotropies \cite{Spergel:2003cb}, \cite{Spergel:2006hy} as well as the baryon acoustic 
oscillations (BAO) in the Sloan Digital Sky Survey (SDSS) luminous galaxy sample \cite{Eisenstein:2005su}. 
      All usual types of matter with positive pressure generate attractive forces and 
decelerate the expansion of the universe. For that reason, a dark energy component with negative 
pressure was suggested to account for a fluid that drives the current accelerated expansion 
(see \cite{Sahni:1999gb} for reviews).  The simplest explanation of dark energy is provided 
by a cosmological constant, but the scenario is plagued by a severe fine tuning problem associated 
with its energy scale. The vacuum energy density falls below the value predicted by any sensible quantum 
field theory by many orders of magnitude \cite{Weinberg:1988cp}, and it unavoidably leads to the coincidence 
problem, i.e.,``Why are the vacuum and matter energy densities of precisely the same order 
today?" \cite {Steinhardt}. 
More sophisticated models replace the cosmological constant by a dynamical dark energy that may be a scalar field 
(quintessence), tachyon field, phantom field or exotic equations of state. These 
models fit the observational data but it is doubtful that they can solve the coincidence problem 
\cite{Chimento:2000kq},\cite{Chimento:2003iea}.  Besides, in a viable dark energy scenario we require that the 
energy density of the dark fluid remains subdominant during radiation and matter dominant eras and that it 
becomes important only at late times to account for the current acceleration of the Universe. 
 
In several papers, it has been proposed that dark matter and dark energy are coupled and do not 
evolve separately \cite{Binder:2006yc},\cite{Amendola:1999er}. The coupling between matter 
and quintessence is either motivated by high energy particle physics considerations \cite{Amendola:1999er} 
or is constructed by requiring the final matter to dark energy ratio to be stable against 
perturbations \cite{Chimento:2003iea},\cite{Zimdahl:2001ar}.  In the same sense, we will use two 
arbitrary interacting fluids in the dark sector, with constant barotropic indexes, 
and will show that there is a 
stable solution for the ratio dark matter-dark energy, which satisfies the expected behavior from the matter 
dominated era until today. We explain in section II the general features of the model for two different interactions and find the stability conditions of the ratio dark matter-dark energy. 
The section III is devoted to obtain the values of parameters that best 
fit the experimental data using the recently published Hubble function $H(z)$ data 
\cite{Simon:2004tf}, extracted from differential ages of passively evolving galaxies. The allowed regions of 
probability $1\sigma$ and $2\sigma$ for the parameters are showed. Also, deceleration parameter, 
densities parameters, ratio dark matter-dark energy, effective equations of state and luminosity distance are exhibited in section IV for the best fit parameters, showing the correct behavior expected for a viable scenario. 
Finally, in section V we 
present our conclusions.

\section{Interacting Dark Model} 
 
Let us consider a Universe modeled by a mixture of three constituents, namely, baryons, 
dark matter and dark energy. The Friedmann and energy conservation 
equations read 
 \ben\n{1} 
 3H^{2}=\rho_1+\rho_2+\rho_3,\\\n{1a} 
\dot\rho_1+\dot\rho_2+3H(\rho_1+p_1+\rho_2+p_2)=0,\\\n{1b} 
 \dot\rho_3+3H(\rho_3+p_3)=0. 
 \een 
respectively. The subindexes 1 and 2 refer to dark matter and dark energy, respectively, 
whereas the subindex 3  to baryons. The baryonic  matter is supposed to be a 
noninteracting component so that it decouples from equation (\ref{1a}). Moreover, by considering the baryons as pressureless $(p_3=0)$, the integration of equation (\ref{1b}) gives $\rho_3=\rho_3^0\left({a_0/a}\right)^3$, where $\rho_3^0$ and $a_0$ represent the present values of the energy density and cosmic scale factor, respectively. 
 
 Hence, equation (\ref{1a}) express the interaction between the dark matter and dark energy 
 allowing the mutual exchange of energy and momentum. Consequently, there will be no 
 local energy-momentum conservation for the fluids separately. 
 However, we can decoupled equation (\ref{1a}) into two "effective equations of conservation" as follows 
 \ben\n{5} 
 \dot\rho_1+3H\ga_1^e\rho_1=0,\\\n{5a} 
 \dot\rho_2+3H\ga_2^e\rho_2=0. 
 \een 
 In the above equations it was introduced effective barotropic indexes $\ga_i^e \,(i=1,2)$ given by 
 \ben\n{6} 
 \ga_1^e=\ga_1+\frac{\ga_2}{r}+\frac{\dot\rho_2}{3H\rho_1},\\ 
 \ga_2^e=\ga_2+\ga_1 r+\frac{\dot\rho_1}{3H\rho_2}, 
 \een 
 where $r=\rho_1/\rho_2$ is the ratio between the energy densities and 
 $\gamma_i \, (i=1,2)$ represent  barotropic indexes of the equations of state 
 $p_i=(\gamma_i-1)\rho_i$. 
 
 In addition we get the relationship 
 \be\n{7} 
 (\ga_1^e-\ga_1)r+(\ga_2^e-\ga_2)=0, 
 \ee 
 and a dynamical equation for the ratio $r$, namely 
 \be\n{8} 
 \dot r=-3Hr\D\ga^e,  \qquad \D\ga^e=\ga_1^e-\ga_2^e. 
 \ee 
 
 We assume that the effective barotropic index of the dark energy is given by $\ga_2^e=\ga_2-F(r)$, where $F(r)$ is a function which depends on the ratio $r$, so that we get 
 \be\n{9} 
 \D\ga^e=\D\ga+F(r)\left(1+\frac{1}{r}\right), 
 \qquad \D\ga=\ga_1-\ga_2. 
 \ee 
 
 By taking into account the above representation for $\gamma_2^e$, Eqs. (\ref{5}) and  (\ref{5a}) 
 can be rewritten as 
 \ben\n{5b} 
 \dot\rho_1+3H\rho_1\gamma_1=-3H\rho_2F(r),\\\n{5c} 
 \dot\rho_2+3H\rho_2\gamma_2=3H\rho_2F(r). 
 \een 
 Now it is possible to identify the right-hand side of each  above equations as the energy transfer 
 between dark matter and dark energy. 
 
Now we make a detailed study of the general dynamics of the 
density ratio $r$ as given by Eq. (\ref{8}) along with Eq. (\ref{9}). To this end we assume that the constant solutions $r=r_s$, represents a stationary stage of the Universe. Then, according to Eq. (\ref{8}), it means that $\D\ga^e(r_s)=0$. Then, the constant solutions $r_s$ will be asymptotically stable whether $(d\D\ga^e/dr)_{r=r_s}>0$. Expressing this condition in terms of Eq. (\ref{9}),  we obtain the stability condition 
\be 
\n{cf} 
r_s\left(1+r_s\right)\left(\frac{dF}{dr}\right)_{r=r_s}-F(r_s)>0, 
\ee 
where we have assumed that the barotropic indexes $\ga_i$ of the two fluid are constants. Note that the simplest choice, a negative constant $F$, satisfies the latter condition. This fact, induce us to investigate the $F$ constant case. So, in the next sections we shall analyze two different choices for the function $F(r)$, namely, one refers to a constant and another being a variable. 
 
\subsection{$F(r)$= constant$<0$} 
 
If we choose the barotropic indexes  $\ga_1$ and  $\ga_2$ as constants and the function $F(r)$ as 
\be\n{F} 
F(r)=-\frac{r_\infty}{1+r_\infty}\D\ga, 
\ee 
where $r_\infty$ is a constant value of the ratio between the energy densities at infinity, 
then it is possible to integrate the equation (\ref{8}) for $r$ along with the effective conservation equations (\ref{5}) and (\ref{5a}) for the energy densities, yielding 
\ben\n{10} 
\rho_1=\rho_2^0\left\{r_\infty+(r_0-r_\infty)\left(\frac{a_0}{a}\right)^{3\al}\right\} 
\left(\frac{a_0}{a}\right)^\beta,\\\n{11} 
\rho_2=\rho_2^0\left(\frac{a_0}{a}\right)^\beta, 
\een 
where $a_0$ and $r_0$ are the present values of the cosmic scale factor of the ratio of 
the energy densities, whereas 
\be 
\n{alpha} 
\al=\frac{\D\ga}{1+r_\infty}, \qquad \beta=3\frac{r_\infty\ga_1+\ga_2}{1+r_\infty}. 
\ee 
 
As usual we introduce the red-shift $z=1/a-1$ and the density parameters at the present value $z_0=0$, namely 
 \be 
 \Omega_1^0=r_0\Omega_2^0,\quad 
 \Omega_2^0={\rho_2^0\over 3 H_0^2},\quad 
 \Omega_3^0={\rho_3^0\over 3 H_0^2}, 
 \ee 
with $\Omega_1^0+\Omega_2^0+\Omega_3^0=1$. Once the explicit dependence of the energy densities are known in terms of the red-shift 
 we can rewrite the Friedmann equation (\ref{1}), thanks to (\ref{10}) and (\ref{11}), as 
 \ben\nonumber 
 H^2=H_0^2\left\{\Omega_2^0(1+r_\infty)(1+z)^\beta+\Omega_3^0(1+z)^3 
 \right. 
 \\ 
 \left. 
 +[1-\Omega_3^0-(1+r_\infty)\Omega_2^0](1+z)^{3\gamma_1}\right\}, 
 \een 
 which gives the dependence of the Hubble parameter $H$ with respect to the red-shift.

\subsection{Variable $F(r)$} 

 One  possible choice for the function $F(r)$ is 
 \be 
 \n{fs} 
 F(r)=-{(1-r)r_\infty^2\over r(1-r_\infty^2)}\Delta\gamma, 
 \ee 
 where $r_\infty$ has the same meaning as in the previous case. Note that the above 
 chosen function  reduces to (\ref{F}) in the limit $r=r_\infty$. 
 With this restriction the latter function satisfies the stability condition (\ref{cf}). 
  So that, the stationary solution of Eq. (\ref{8}), corresponding to the 
  function (\ref{fs}), is asymptotically stable. 
 
In this case the solution of Eqs. (\ref{5}) and (\ref{5a}) are given by 
 
 \[ 
 \rho_1=\rho_2^0\sqrt{r_\infty^2+(r_0^2-r_\infty^2)\left({a_0\over a}\right)^\nu} 
 \] 
 \be 
 \times\left({a_0\over a}\right)^{3\epsilon} 
 \left[{(1-r/r_\infty)(1+r_0/r_\infty)\over 
 (1-r_0/r_\infty)(1+r/r_\infty)}\right]^{r_\infty\over2}, 
 \ee 
 
 \be 
 \rho_2=\rho_2^0\left({a_0\over a}\right)^{3\epsilon} 
 \left[{(1-r/r_\infty)(1+r_0/r_\infty)\over 
 (1-r_0/r_\infty)(1+r/r_\infty)}\right]^{r_\infty\over2}, 
 \ee 
where 
\be 
\nu={6\Delta\gamma\over 1-r_\infty^2}, \qquad \epsilon=\gamma_2-{\Delta\gamma r_\infty^2\over 1-r_\infty^2}. 
\ee 
Furthermore, the Hubble parameter $H$ can  be expressed as functions of the red-shift, yielding 
 \[ 
 {H^2\over H^2_0}=\Omega_2^0(1+z)^{3\epsilon} 
 \Biggl[{(1-r/r_\infty)(1+r_0/r_\infty)\over 
 (1-r_0/r_\infty)(1+r/r_\infty)}\Biggr]^{r_\infty\over2} 
 \] 
 \be 
  \times\Biggl\{1+ \sqrt{r_\infty^2+(r_0^2-r_\infty^2)(1+z)^{\nu}} 
 \Biggr\}+\Omega_3^0(1+z)^3. 
 \ee 
 
  Here it is  interesting to call attention that  the condition $r_\infty=0$ 
 leads to a mixture of noninteracting fluids with constant barotropic indexes for both cases analyzed above. 
 
\section{Cosmological Constraints} 

\begin{figure} 
 \begin{center} 
\includegraphics[width=6.0cm,height=7.0cm]{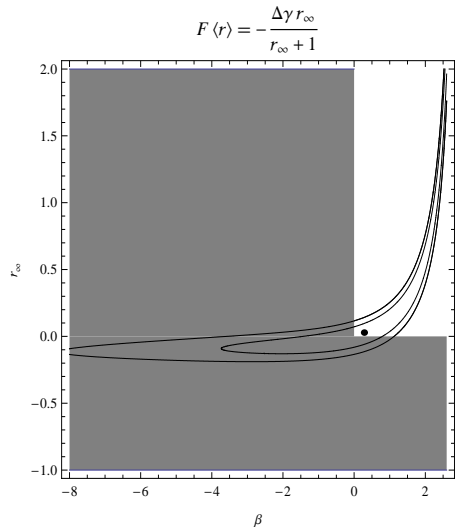} 
\caption{Constant $F$: $\beta $ vs. $ r_\infty$.} 
\end{center}
\end{figure}

The aim of this section is to determine the parameters $(r_\infty, \beta)$ for the constant $F(r)$ case and 
$(r_\infty, \gamma_2)$ for the variable $F(r)$ one.  These parameters, denoted generically 
by $a_i$, are established   through the recently published Hubble 
parameter $H(z)$ data 
\cite{Simon:2004tf}, extracted from 
differential ages of passively evolving galaxies. This $H(z)$ function is a very interesting one, because in contrast to standard candle luminosity distances or standard ruler angular diameter distances, it is not integrated over.  The Hubble parameter depends on the differential 
age of the Universe as a function of $z$ in the form 
$H(z) = -(1 + z)^{-1}dz/dt$, and it can be measured directly  through a determination of $dz/dt$. In the procedure of calculating the differential ages, Simon et al. \cite{Simon:2004tf} have employed the new released Gemini Deep Deep Survey \cite{Abraham:2004ra} and archival data \cite{Treu:2001hq}, \cite{Nolan:2003bt} to determine the 9 numerical values of $H(z)$ in the range $0 < z < 1.8$.

The probability distribution for the parameters 
$a_1,a_2$ is (see e.g. \cite{Press1994}) 
\be 
P(a_1,a_2) = \mathcal{N} e^{-\chi^2(a_1,a_2)/2}, 
\ee 
where $\mathcal{N}$ is a normalization constant. Hence, we minimize the $\chi^2$ function 
\be 
\chi^2(a_1,a_2)= \sum_{i=1}^{9}\frac{[H_{th}(a_1 
,a_2; z_i) - H_{ob}(z_i)]^2}{\sigma(z_i) ^2} 
\ee 
 where $H_{ob}(z_i)$ is the observed value of $H$ at the red-shift $z_i$, 
 $\sigma(z_i)$ is the corresponding $1\sigma$ uncertainty, and the summation 
 is over the 9 observational $H(z_i)$ data points at red-shift $z_i$ \cite{Simon:2004tf}. 
We adopt the priors $H_0=72$ km s$^{-1}$ Mpc$^{-1}$ (mean value of the 
results from the Hubble Space Telescope key project), $ \Omega_1^0=0.25$, 
$\Omega_2^0=0.70$  and $\Omega_3^0 =0.05$ \cite{Freedman:2000cf}. Furthermore, we assume 
a spatially flat Universe with a pressureless (dust)  dark matter, i.e., $\gamma_1=1$. Besides, we take the theoretical expressions for the Hubble parameter $H_{th}$ with parameters $a_1=r_\infty$, $a_2=\beta$ in the $F$ constant case and $a_1=r_\infty$ and  $a_2=\gamma_2$, in the $F$ variable case. 
 
\begin{figure} 
 \begin{center} 
\includegraphics[width=8cm,height=7cm]{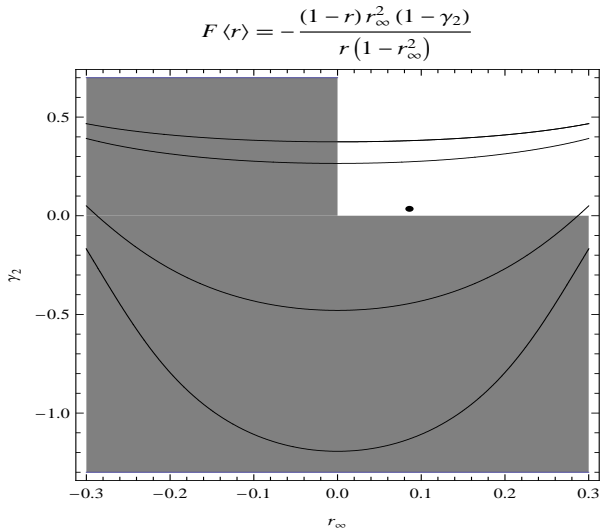} 
\caption{Variable $F$: $r_\infty$ vs. $\gamma_2$.} 
\end{center} 
\end{figure} 
 
The best fit parameters $a_{1bf}$ and $a_{2bf}$ are obtained so that $\chi_{min}^2(a_{1bf},a_{2bf})$ 
corresponds to the minimum of $\chi^2(a_1,a_2)$. If $\chi_{min}^2(a_{1bf},a_{2bf})/(N - 
n) \leq 1$ the fit is good and the data are consistent with 
the considered model $H(z; a_1, a_2)$. Here, $N$ is the range of data set used and $n$ is the number of parameters \cite{Press1994}. 
 
The variable $\chi^2$ is a random variable in the sense that it depends 
on the random data set used. Its probability distribution 
is a $\chi^2$ distribution for $N-n$ degrees of freedom. In our cases, this implies that 68\% of the random data sets will give a $\chi^2$ such that 
\be 
\chi^2(a_1, a_2) - \chi^2(a_{1bf},a_{2bf}) \leq 2.3. 
\ee 
 This equation defines a closed elliptical 
curve around $a_{1bf}$ and $a_{2bf}$ in the bi-dimensional parameter 
space. The corresponding $1\sigma$ range of the parameter $a_i$ is 
the range of $a_i$ for points contained within this elliptical 
curve. Similarly, it can be shown that 95.4\% of the 
random data sets will give a $\chi^2$ such that 
\be 
\chi^2(a_1, a_2) - \chi^2(a_{1bf},a_{2bf}) \leq 6.17. 
\ee 
Again, this last equation defines an elliptical curve in parameter space and the corresponding $2\sigma$ range of the each parameter $a_i$ is the range of $a_i$ for points contained within this elliptical curve.

The local minimum in the case of $F$ constant is $\chi^2 = 9.02445$  for $r_\infty = 0.0230767 $ and $\beta = 0.302457$. Hence, it follows from eq. (\ref{alpha}) with $\gamma_1 = 1$ that  $\gamma_2 = 0.0800689$ and $\alpha = 0.899181$. In the $F$ variable case, the minimum is  $\chi^2 = 9.03738$ for $r_\infty = 0.0851298$ and $\gamma_2 = 0.0328392$. 
 
 In Figs. 1 and 2 we have plotted the confidence regions, that is, the sections of the elliptical curves explained above, in the $\beta $ vs. $ r_\infty$ plane for a constant $F$ and in the $r_\infty$ vs. $\gamma_2$ plane for a variable $F$, respectively. The points inside the inner ellipses or between both ellipses identify the true values of parameters with 68.3\% or 95.4\%  of probability, respectively. Also, in both figures, the best fit value for each model is represented by a dot   and the shady zone means the forbidden negative values.

\section{Cosmological Solutions} 
 
 \begin{figure} 
 \begin{center} 
 \includegraphics[width=7cm]{ff3.eps}\hskip0.25cm 
 \includegraphics[width=7cm]{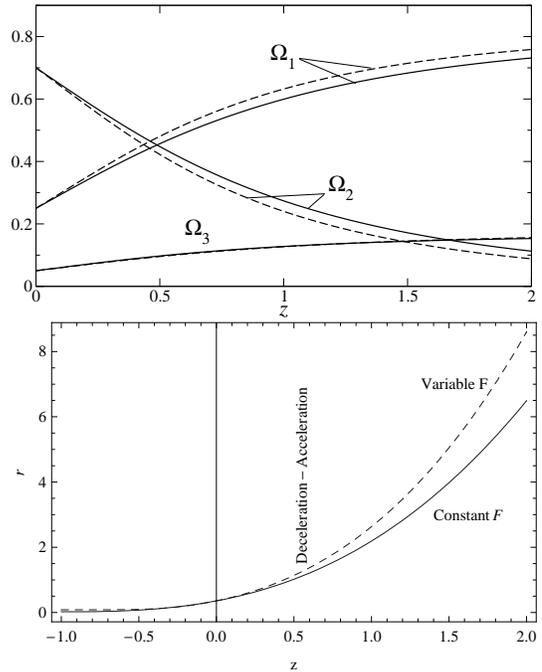} 
 \caption{Top: density parameters as functions of the red-shift; bottom: 
 ratio between dark matter and dark energy as function of the red-shift. 
  Straight lines -- constant $F$; dashed lines -- variable $F$.} 
 \end{center} 
 \end{figure}

In this section we search for cosmological solutions of the model proposed in previous sections. First we analyze 
the density parameters which are plotted as functions of the red-shift in the top of Fig. 3,  the straight lines corresponding to the constant $F(r)$, whereas the dashed lines to the variable $F(r)$. One can infer from this figure that the energy transfer from the dark energy to dark matter is more pronounced for the variable case, since for this case the growth of the density parameter of the dark matter  and the corresponding decay of the dark energy with the red-shift are more pronounced than those for a constant $F(r)$. This behavior can also be verified from the bottom of Fig. 3 which represents the evolution of the ratio of the two energy densities $r=\rho_1/\rho_2$ with the red-shift. This last figure also shows that in the future, i.e., for negative values of the red-shift, there is no difference between the two cases, since both tend to a small values, indicating a predominance of the dark energy in the future. 
 
\begin{figure} 
 \begin{center} 
 \includegraphics[width=7cm]{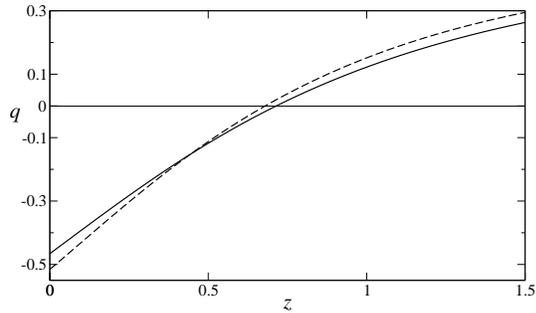} 
 \caption{Deceleration parameter as function of the red-shift. Straight line -- constant $F$; dashed line -- variable $F$.} 
 \end{center} 
 \end{figure} 
 
In Fig. 4 the deceleration parameter $q=1/2+3p/2\rho$  is plotted  for the two cases. The 
present values of the deceleration parameter $q(0)$ and the value for the red-shift $z_t$ where  the transition 
from a decelerated to an accelerated regime occur are : (i) $q(0)\approx-0.47$ and $z_t\approx0.72$ for the constant $F$  and 
(ii) $q(0)\approx-0.52$ and $z_t\approx0.68$ for the variable $F$. These values are of the same order of magnitude of the experimental values: $q(0)=-0.74\pm0.18$ (see~\cite{Vir}) and $z_t=0.46\pm0.13$ (see~\cite{riess}). 
 
The equations of state for dark matter and dark energy can be written in terms of  the effective parameters 
$w_1^e=\gamma_1^e-1$ and $w_2^e=\gamma_2^e-1$ as 
 \be 
 p_1=w_1^e\rho_1, \qquad 
 p_2=w_2^e\rho_2, 
 \ee 
 respectively. In Fig. 5 we have plotted the effective parameters $w_1^e$ and $w_2^e$ as functions of the red-shift $z$. For the variable case, one can observe that in the red-shift range $-1\leq z\leq2$  the effective parameters assume the values between: $-0.89\lesssim w_1^e\lesssim 7.2\times 10^{-4}$ and $-0.89\lesssim w_2^e \lesssim-0.97$. Hence, in this model the effective equation of state of the dark matter constituent behaves practically as a pressureless (dust) fluid for $z>0$ and as quintessence for $z<0$. The dark energy in the same range behaves always as  quintessence.

 \begin{figure} 
 \begin{center} 
 \includegraphics[width=7cm]{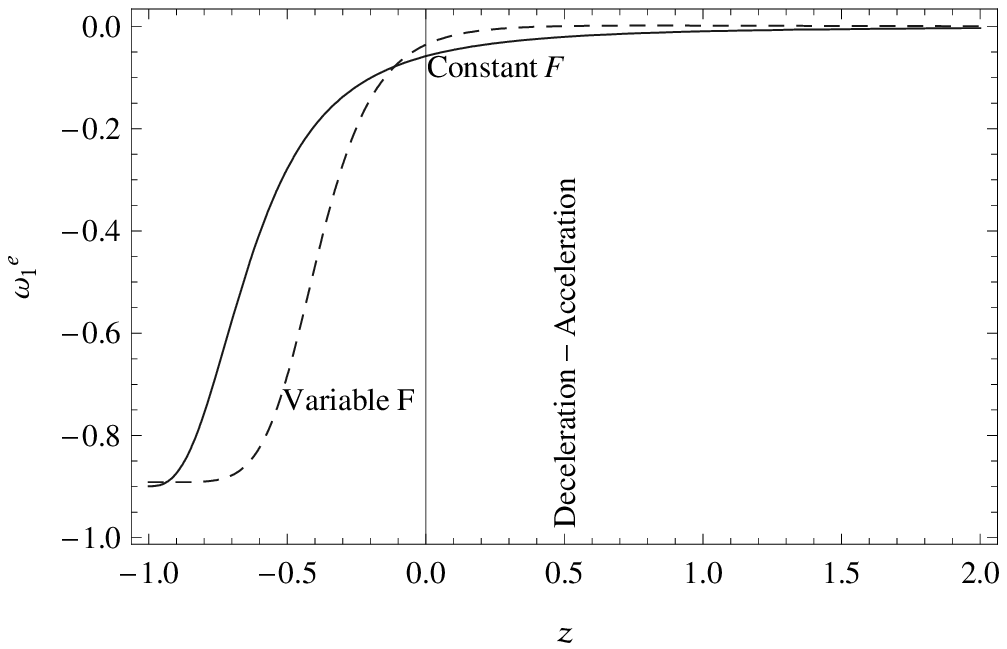}\hskip0.5cm 
 \includegraphics[width=7cm]{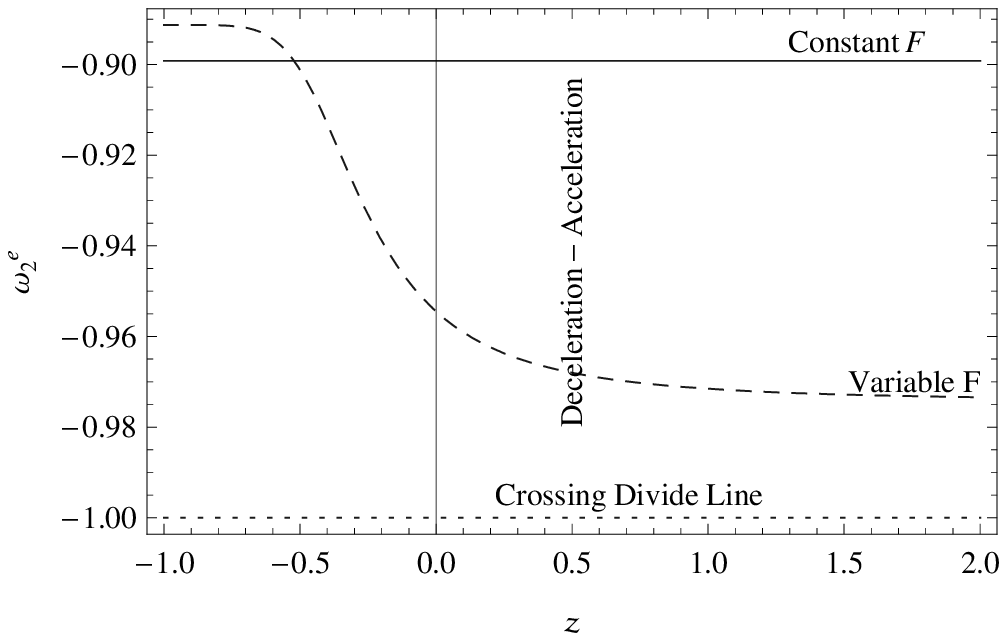} 
 \caption{$w_1^e$ and $w_2^e$ as functions of the red-shift $z$. Top: $w_1^e$;  bottom: $w_2^e$.} 
 \end{center} 
 \end{figure} 
 \begin{figure} 
 \begin{center} 
 \includegraphics[width=7cm]{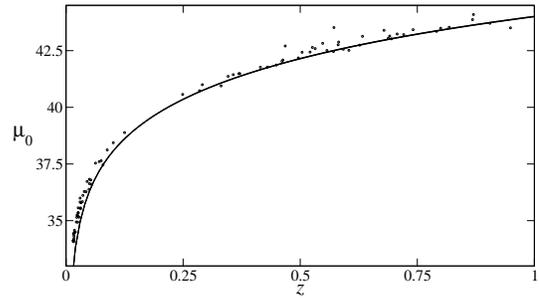} 
 \caption{$\mu_0$ as function of the red-shift $z$.} 
 \end{center} 
 \end{figure} 
 
 In Fig. 6 it is represented the difference between the apparent magnitude $m$ 
 and the absolute magnitude $M$ of a source, denoted by $\mu_0$ and whose expression is 
 \be 
 \mu_0=m-M=5\log \left\{(1+z)\int_0^z{dz'\over 
 H(z')}\right\}+25, 
 \ee 
 with the quantity between braces representing the luminosity distance in Mpc. The 
 circles in Fig. 6 are  experimental data for super-novae of type Ia taken from the work \cite{14}. It is interesting to note that all $\mu_0$ -- curves  corresponding to  the two models analyzed in the present work and the  $\Lambda$CDM model practically coincide, the only small difference between them is for higher values of the red-shift. Furthermore, there is a good fitting of these curves with the experimental data.

\section{Conclusions} 
To sum up: in this work a cosmological model for the present Universe was proposed whose constituents were 
  the non-interacting baryonic matter and 
interacting dark components. The interaction between the dark energy and dark matter 
was related with their effective barotropic indexes, which were considered as functions of the ratio between their energy densities. Furthermore, the parameters of the functions were adjusted by considering best fits to Hubble function data. 
Two functions were analyzed, namely, a constant  and a variable one. 
It was shown: (i) for the proposed functions the ratio $r$ is asymptotically stable; (ii) the energy transfer from the dark energy 
to the dark matter is more efficient for the variable case; (iii) for both functions the dark energy density predominates in the future; (iv) for both functions the present values of the deceleration parameter and the values of the transition red-shift -- where the regime changes from a decelerated to an accelerated -- are of the same order as the experimental values; (v) for the variable function, the effective ratio between the pressures and energy densities for the dark components 
indicates that the dark energy behaves as quintessence whereas the dark matter as practically a pressureless fluid (dust) when $z>0$;  (vi) the behavior of the parameter $\mu_0$ -- related with the luminosity distance -- with the red-shift does not have a sensible difference for the two functions and practically coincide with the one of the $\Lambda$CDM model, indicating a good fit with the experimental values.

\section*{Acknowledgments} 
 
The authors acknowledge the partial 
support under project 24/07 of the  agreement SECYT (Argentina) and CAPES 117/07 (Brazil). 
LPC thanks the University of Buenos Aires for partial support under 
project X224, and the Consejo Nacional de Investigaciones 
Cient\'{\i}ficas y T\'ecnicas under project 5169. GMK acknowledges the support by 
Conselho Nacional de Desenvolvimento Cient\'\i fico e Tecnol\'ogico (CNPq).


\begin{thebibliography}{99} 
 
\bibitem{Riess:1998cb} 
  A.~G.~Riess { et al.} 
  Astron.\ J.\  {\bf 116}, 1009 (1998); 
  A.~G.~Riess { et al.}, 
  Astron.\ J.\  {\bf 117}, 707 (1999). 
 
 
 
 
\bibitem{Perlmutter:1998np} 
  S.~Perlmutter { et al.}  
  Astrophys.\ J.\  {\bf 517}, 565 (1999). 
 
 
\bibitem{Astier:2005qq} 
  P.~Astier { et al.}  
  Astron.\ Astrophys.\  {\bf 447}, 31 (2006). 
 
 
\bibitem{Spergel:2003cb} 
  D.~N.~Spergel {et al.}  
  Astrophys.\ J.\ Suppl.\  {\bf 148}, 175 (2003). 
 
\bibitem{Spergel:2006hy} 
  D.~N.~Spergel { et al.}  
  Astrophys.\ J.\ Suppl.\  {\bf 170}, 377 (2007). 
 
 
\bibitem{Eisenstein:2005su} 
  D.~J.~Eisenstein { et al.}  
  Astrophys.\ J.\  {\bf 633}, 560 (2005). 
 
 
 
\bibitem{Sahni:1999gb} 
  V.~Sahni and A.~A.~Starobinsky, 
  Int.\ J.\ Mod.\ Phys.\  D {\bf 9}, 373 (2000); 
  V.~Sahni, in {\it The Physics of the Early Universe,} 
  Ed. E. Papantonopoulos, 
  Lect.\ Notes Phys.\  {\bf 653} (Springer, Berlin, 2005); 
  S.~M.~Carroll, 
  Living Rev.\ Rel.\  {\bf 4}, 1 (2001); 
  T.~Padmanabhan, 
  Phys.\ Rept.\  {\bf 380}, 235 (2003); 
  P.~J.~E.~Peebles and B.~Ratra, 
  Rev.\ Mod.\ Phys.\  {\bf 75}, 559 (2003); 
  E.~J.~Copeland, M.~Sami and S.~Tsujikawa, 
  Int.\ J.\ Mod.\ Phys.\  D {\bf 15}, 1753 (2006). 
 
\bibitem{Weinberg:1988cp} 
  S.~Weinberg, 
  Rev.\ Mod.\ Phys.\  {\bf 61}, 1 (1989). 
 
\bibitem{Steinhardt} 
 P.J. Steinhardt, in {\it Critical Problems in Physics}, 
  Ed.  V.L. Fitch, D.R. Marlow and M. A. E. Dementi 
 (Princeton University Press, Princeton,  1997). 
 
 
\bibitem{Chimento:2000kq} 
  L.~P.~Chimento, A.~S.~Jakubi and D.~Pavon, 
  Phys.\ Rev.\  D {\bf 62}, 063508 (2000). 
 
\bibitem{Chimento:2003iea} 
  L.~P.~Chimento, A.~S.~Jakubi, D.~Pavon and W.~Zimdahl, 
  Phys.\ Rev.\  D {\bf 67}, 083513 (2003). 
 
 
 
\bibitem{Binder:2006yc} 
  J.~B.~Binder and G.~M.~Kremer, 
  Gen.\ Rel.\ Grav.\  {\bf 38}, 857 (2006); 
  D.~Tocchini-Valentini and L.~Amendola, 
  Phys.\ Rev.\  D {\bf 65}, 063508 (2002); 
  G.~R.~Farrar and P.~J.~E.~Peebles, 
  Astrophys.\ J.\  {\bf 604}, 1 (2004); 
  G.~M.~Kremer, 
  Gen.\ Rel.\ Grav.\  {\bf 39}, 965-972 (2007); 
  G.~Huey and B.~D.~Wandelt, 
  Phys.\ Rev.\  D {\bf 74}, 023519 (2006); 
  G.~Mangano, G.~Miele and V.~Pettorino, 
  Mod.\ Phys.\ Lett.\  A {\bf 18}, 831 (2003); 
  L.~P.~Chimento and M.~Forte, 
  [arXiv:0706.4142 [astro-ph]; 
  R.~G.~Cai and A.~Wang, 
  JCAP {\bf 0503}, 002 (2005). 
 
\bibitem{Amendola:1999er} 
  L.~Amendola, 
  Phys.\ Rev.\  D {\bf 62}, 043511 (2000); 
  L.~Amendola and D.~Tocchini-Valentini, 
  Phys.\ Rev.\  D {\bf 64}, 043509 (2001); 
  L.~Amendola and D.~Tocchini-Valentini, 
  Phys.\ Rev.\  D {\bf 66}, 043528 (2002); 
  L.~Amendola, C.~Quercellini, D.~Tocchini-Valentini and A.~Pasqui, 
  Astrophys.\ J.\  {\bf 583}, L53 (2003). 
 
\bibitem{Zimdahl:2001ar} 
  W.~Zimdahl and D.~Pavon, 
  Phys.\ Lett.\  B {\bf 521}, 133 (2001). 
 
\bibitem{Simon:2004tf} 
  J.~Simon, L.~Verde and R.~Jimenez, 
  Phys.\ Rev.\  D {\bf 71}, 123001 (2005). 
 
 
 
 
\bibitem{Abraham:2004ra} 
  R.~G.~Abraham { et al.}, 
  Astron.\ J.\  {\bf 127}, 2455 (2004). 
 
 
 
 
 
\bibitem{Treu:2001hq} 
  T.~Treu, M.~Stiavelli, P.~Moller, S.~Casertano and G.~Bertin, 
  Mon.\ Not.\ Roy.\ Astron.\ Soc.\  {\bf 326}, 221 (2001). 
 
\bibitem{Nolan:2003bt} 
  P.~L.~Nolan, W.~F.~Tompkins, I.~A.~Grenier and P.~F.~Michelson, 
  Astrophys.\ J.\  {\bf 597}, 615 (2003). 
 
\bibitem{Press1994} 
W.H. Press et al., {\it Numerical Recipes} (Cambridge University 
Press, Cambridge, 1997). 
 
 
 
\bibitem{Freedman:2000cf} 
  W.~L.~Freedman {\it et al.}, 
  Astrophys.\ J.\  {\bf 553}, 47 (2001) 
 
 
 
 
\bibitem{Vir} J. M. Virey et al { Phys. Rev. D} {\bf72},  {061302} (2005). 
 
 
\bibitem{riess} A. G. Riess et al. { Astrophys. J.} {\bf607}, 665 (2004). 
 
\bibitem{14}  A. G. Riess et al. { Astrophys. J.} {\bf659} 98 (2007). 
 
 
\end{thebibliography}
\end{document}